\documentclass[final,1p,times]{elsarticle} 

\usepackage{graphicx}
\usepackage{amssymb} 
\usepackage{amsthm} 
\usepackage{lineno}

\journal{Nuclear Physics A} 

\begin{document}

\begin{frontmatter} 

\title{MUSIC with the UrQMD Afterburner}


\author[auth1]{Sangwook Ryu}
\author[auth1]{Sangyong Jeon}
\author[auth1]{Charles Gale}
\author[auth2]{Bj\"orn Schenke}
\author[auth1,auth3]{Clint Young}

\address[auth1]{McGill University, Montreal, QC, Canada}
\address[auth2]{Brookhaven National Laboratory, Upton, NY, USA}
\address[auth3]{University of Minnesota, Minneaplis, MN, USA}

\begin{abstract} 
 As RHIC is entering the precision measurement era and the LHC is producing
 a copious amount of new data, the role of 3+1D event-by-event viscous
 hydrodynamics is more important than ever to understand the bulk data
 as well as providing the background for hard probes. For more meaningful
 comparison with the experimental data, it is also important that
 hydrodynamics be coupled to the hadronic afterburner.
 In this proceeding we report on preliminary results of
 coupling MUSIC with UrQMD.
\end{abstract} 

\end{frontmatter} 


\section{Introduction}

Full understanding of 
Quark-Gluon Plasma (QGP) in relativistic heavy ion collisions
requires a wide range of theoretical tool sets.
There is little doubt that the initial condition of the nuclei before
the collision as well as the gluon dynamics right after the collision is
governed by classical Yang-Mills dynamics \cite{Gale:2012qm,Gale:2012rq}.
Less than $1\,\hbox{fm}/c$ after the collision, the system
reaches the local equilibrium stage and hydrodynamic expansion begins.
How the system reaches the local equilibrium state so quickly is still hotly
debated, but progress is being made \cite{Kurkela:2011ti}.
In the initial stage of the hydrodynamic expansion, the system is in the QGP
phase. This phase is so dense that hydrodynamics must be valid.
As the system expands and cools, it enters the hadronic phase. 
On the one hand, 
this system should still admit a hydrodynamic description
via the hadronic equation of state. On the other hand, 
as the system becomes more dilute, hadronic kinetic theory models 
should become more accurate.
Therefore, there should be a range of temperatures where both descriptions
are valid. This enables us to switch from one description to another
within a range of temperature slightly below the critical temperature.
In this short proceeding, we report on our effort to deal with 
this transition between
the hydrodynamics phase and the hadronic kinetic theory phase
by coupling 3+1D event-by-event viscous hydrodynamics (MUSIC)
and a hadronic quantum molecular dynamics model (UrQMD).
The initial Glasma phase and its coupling to the hydrodynamic phase 
is dealt with in Refs.\cite{Gale:2012qm,Gale:2012rq}.

\section{3+1D Event-by-Event Viscous Hydrodynamics}

Hydrodynamics was first applied to particle collisions by 
Belenkij and Landau in 1956~\cite{Belenkij:1956cd}.
In more recent times, boost-invariant 2+1D ideal hydrodynamics 
have enjoyed much success in heavy ion phenomenology
\cite{Kolb:2003dz,Huovinen:2003fa}.
Detailed data from RHIC experiments, however, eventually
made it clear that
it is necessary to go beyond the 2+1D ideal hydrodynamic 
with smooth averaged initial conditions.
For instance, the surprisingly large triangular flow
\cite{Richardson:2012kq,Pandit:2012rm,Abelev:2012di}
can never be reproduced in this way.
More recent efforts have improved this
by including non-zero shear viscosity~\cite{Heinz:2005bw,Baier:2006gy},
3+1D dynamics~\cite{Muroya:1997ee,Hirano:2000eu}, and
fluctuating initial 
conditions \cite{Aguiar:2001ac,Holopainen:2010gz,Pang:2012he}.  
However, none of the above mentioned studies implements all three
improvements over the 2+1D ideal hydrodynamics. Only recently, the first
implementation of 3+1D event-by-event viscous hydrodynamics was made by 
3 of the present authors \cite{Schenke:2010nt,Schenke:2010rr,Schenke:2011tv}.

Our implementation of hydrodynamics (named MUSIC)
uses the hyperbolic $\tau$-$\eta$
coordinate system without assuming invariance in $\eta$.
The viscous tensor $\pi^{\mu\nu}$ is calculated
by using a variant of the Israel-Stewart formalism \cite{Baier:2007ix}.
The algorithm chosen to solve the system of hyperbolic equations is
the Kurganov and Tadmor algorithm in the semi-discrete form
\cite{Kurganov:2000} together with Heun's method. 
This implementation is relatively simple yet very stable.  
Glauber approximation provides fluctuating initial conditions
for the energy density.
For the equation of state, the lattice based
parameterization ``s95p-v1'' from \cite{Huovinen:2009yb}
is the default option although other options are also
available. When freezing out at about $T = 140\,\hbox{MeV}$ without the
afterburner, the Cooper-Frye formula
is used to get the final state hadronic spectrum with the full resonance
decays adapted from Heinz and Kolb's 2-D scheme. 
This Cooper-Frye procedure includes the viscous correction 
\cite{Dion:2011pp}. 

\begin{figure}[t]
\begin{center}
\includegraphics[width=0.45\textwidth,height=4cm]{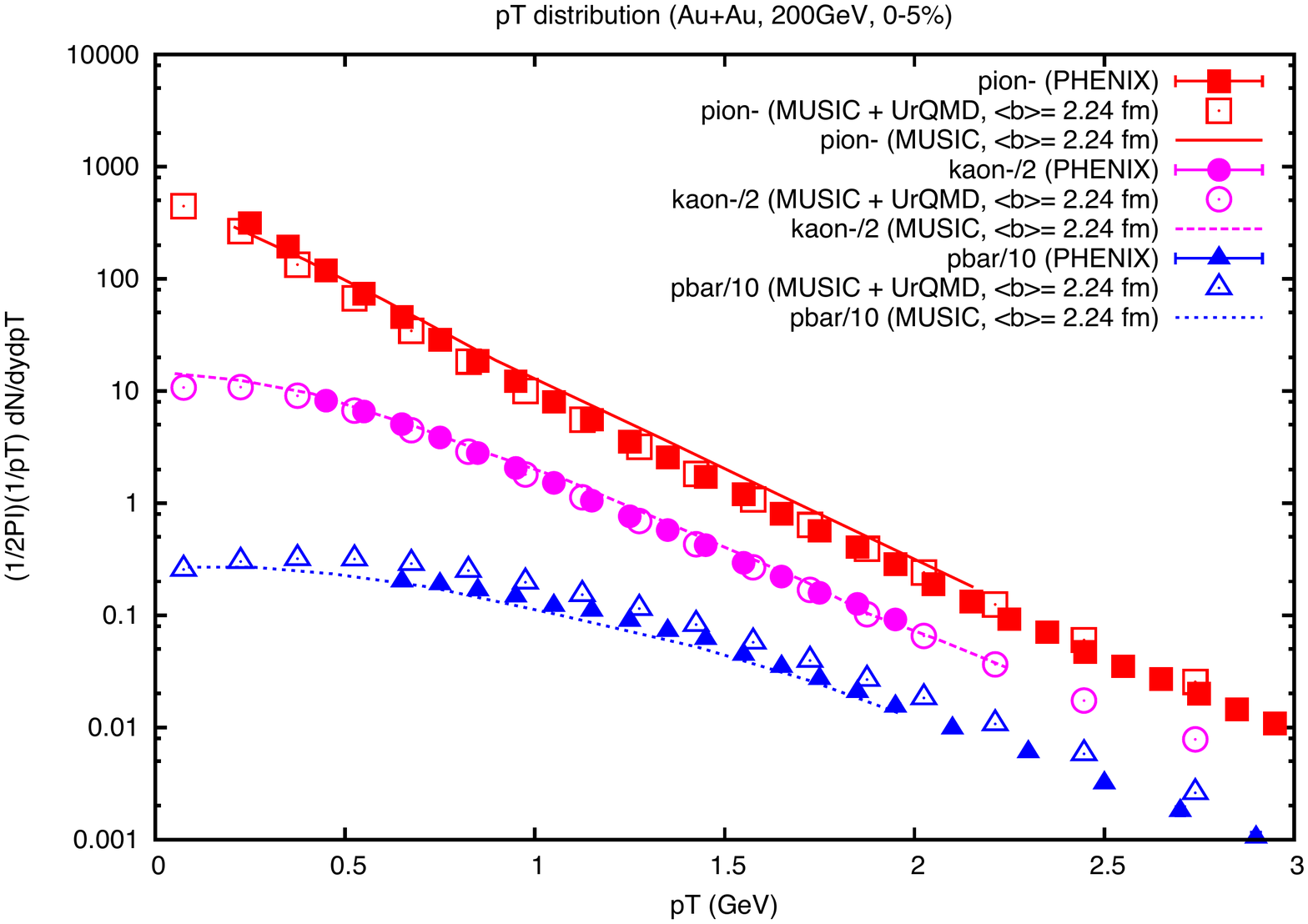}
\includegraphics[width=0.45\textwidth,height=4cm]{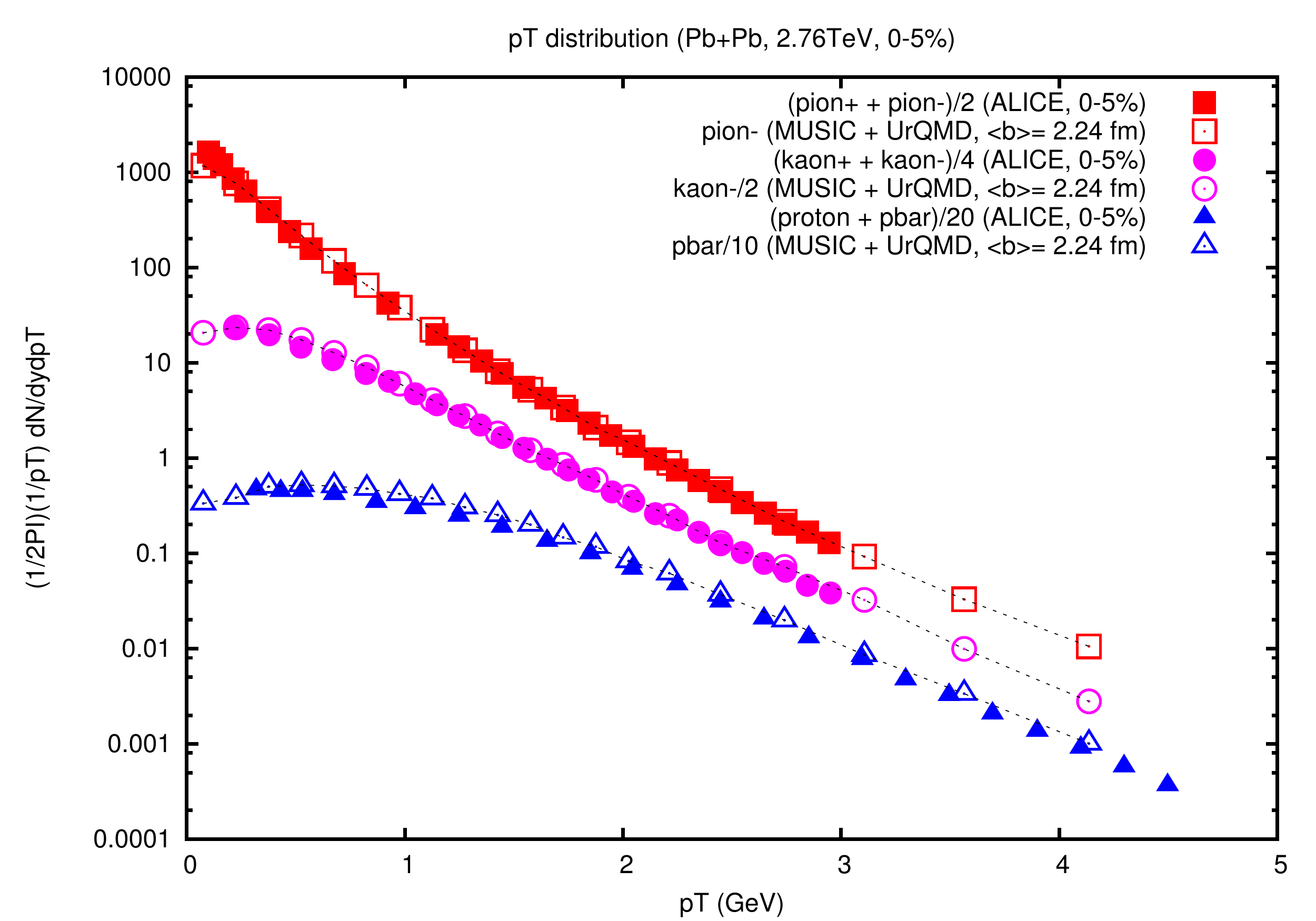}
\end{center}
\caption{
The momentum distribution of RHIC and the LHC 5\,\%
most central collisions
compared to the MUSIC$+$UrQMD calculations. RHIC result is based on 10,000
events (100 UrQMD events on each of 100 hydro events)
events and LHC result is based on 1,000 events 
(10 UrQMD events on each of 100 hydro events).
For RHIC, the results of pure MUSIC events are also shown.
}
\label{fig:dNdpT}
\end{figure}

\begin{figure}[t]
\begin{center}
\includegraphics[width=0.45\textwidth,height=4cm]{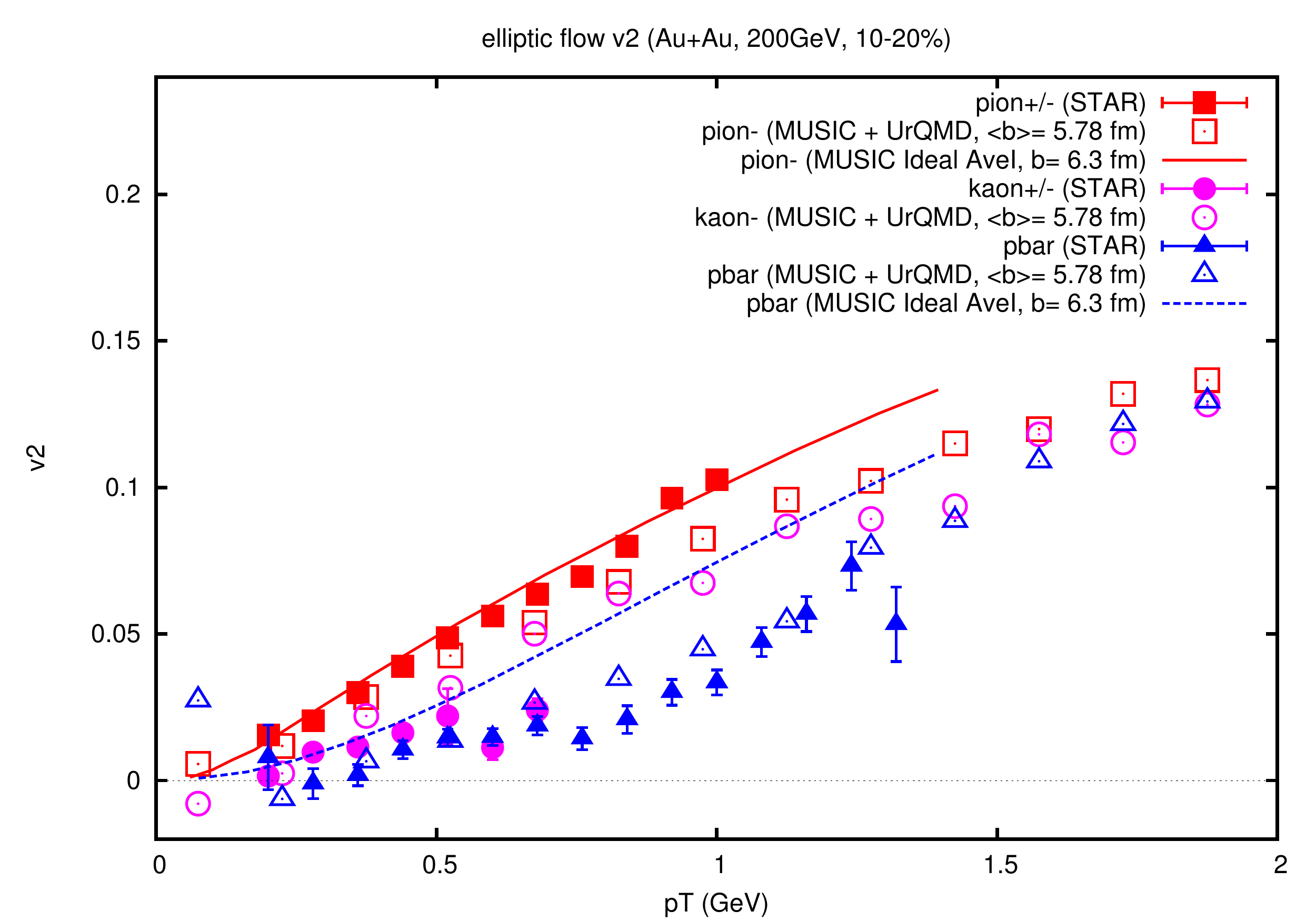}
\includegraphics[width=0.45\textwidth,height=4cm]{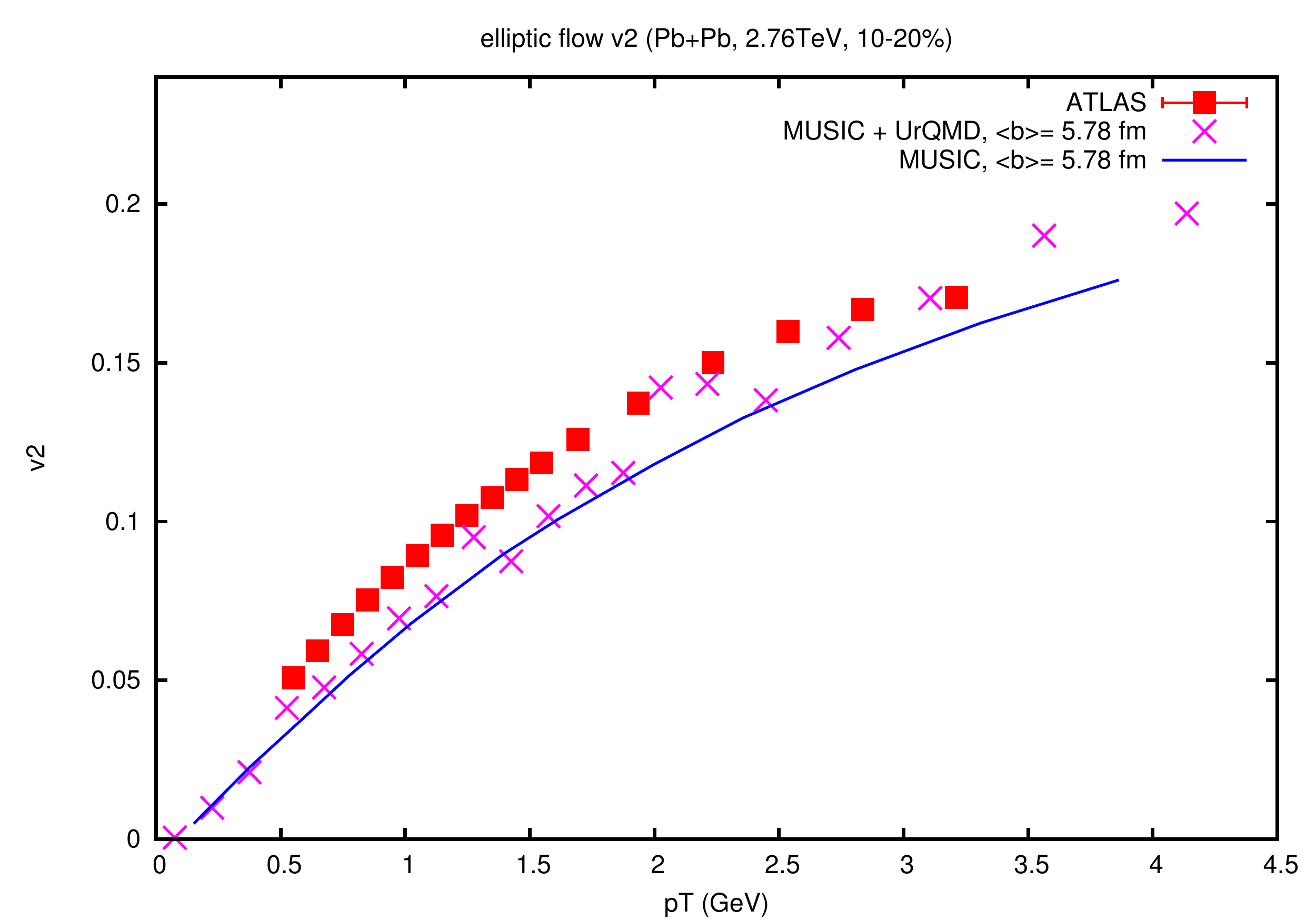}
\end{center}
\caption{
The $p_T$ dependent
$v_2$ from RHIC and the LHC in the $10 - 20\,\%$
centrality class
compared to the MUSIC and MUSIC$+$UrQMD calculations. 
RHIC result is based on 10,000
events (100 UrQMD events on each of 100 hydro events)
events and LHC result is based on 1,000 events 
(10 UrQMD events on each of 100 hydro events).
}
\label{fig:v2}
\end{figure}

\section{UrQMD Afterburner}

UrQMD (Ultrarelativistic Quantum Molecular Dynamics) 
is a hadronic kinetic theory model that has been very
successful in describing heavy ion collisions up to the RHIC energy
\cite{Bass:1998ca,Bleicher:1999xi}. 
Recently, H.~Petersen et~al has updated the UrQMD code to
include hydrodynamics \cite{Petersen:2008dd} and made it 
publicly available.
The hydrodynamics included in the original version (UrQMD v.~3.3p1) 
is a 3+1D ideal hydrodynamic simulation
in the Cartesian coordinate system with
the simple freeze-out surface determined at a constant $t$.
MUSIC on the other hand uses the hyperbolic coordinate system
and generates a fairly complicated iso-thermal freeze-out surface.
We have implemented an
interface between these two systems fully taking into
account the freeze-out hypersurface directions.
(See also Ref.\cite{Huovinen:2012is} for an exact freeze-out-surface finding
algorithm.)
One missing ingredient is the viscous correction of the thermal distribution 
when it is sampled for the subsequent UrQMD run. 
This will be implemented in the 
near future although the viscous correction is not
expected to be large at around $T = 170\,\hbox{MeV}$ 
for moderate values of $p_T$ \cite{Dion:2011pp}.
In this work, we present our preliminary results with
previously tuned values of the hydrodynamics parameters.
More complete analysis including hydro parameter dependencies
will be reported elsewhere.

\begin{figure}[t]
\begin{center}
\includegraphics[width=0.45\textwidth,height=4cm]{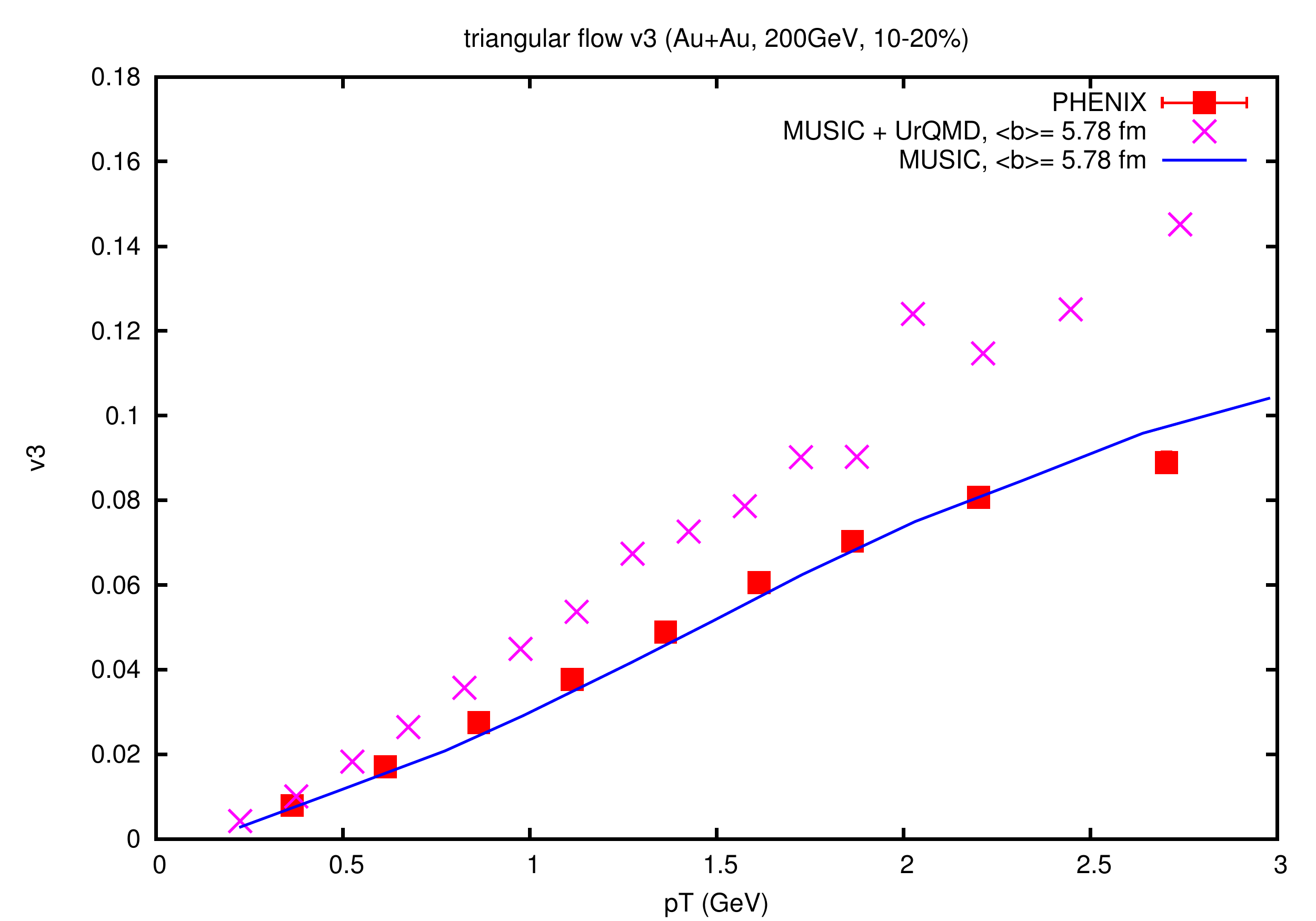}
\includegraphics[width=0.45\textwidth,height=4cm]{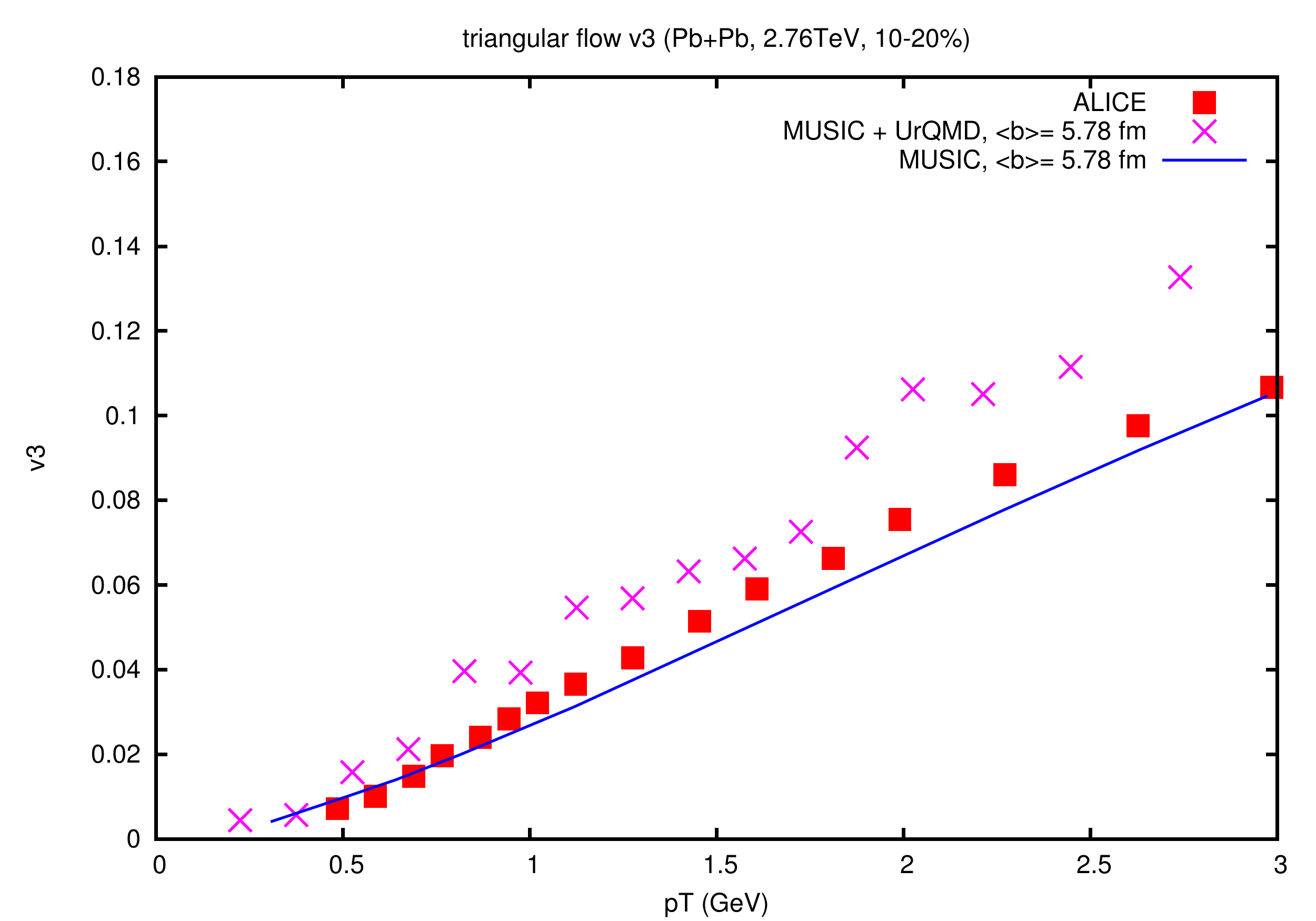}
\end{center}
\caption{
The $p_T$ dependent
triangular flow from RHIC and the LHC in the $10 - 20\,\%$
centrality class
compared to the MUSIC and MUSIC$+$UrQMD calculations. 
RHIC result is based on 10,000
events (100 UrQMD events on each of 100 hydro events)
events and LHC result is based on 1,000 events 
(10 UrQMD events on each of 100 hydro events).
The data points are from Refs.\cite{Richardson:2012kq} (PHENIX)
and \cite{Abelev:2012di} (ALICE).
}
\label{fig:v3}
\end{figure}

\section{Results and Summary}

Three main results are shown in Figures \ref{fig:dNdpT},\ref{fig:v2},
and \ref{fig:v3}.
For RHIC, we have 10,000 events in each centrality bin generated by
running 100 UrQMD events on each of 100 viscous MUSIC events.
For the LHC, we have 1,000 events in each centrality bin
generated by 10 UrQMD events on each of
100 viscous MUSIC events. The transition to UrQMD was made at
$T_{\rm tr} = 170\,\hbox{MeV}$. 
The two panels in Figure \ref{fig:dNdpT} show proof of
principle results. 
The RHIC $p_T$ spectra for $\pi^-$, $K^-$ and $\bar{p}$
are reasonably well produced for the $0 - 5\,\%$ centrality bin. 
For all 3 species, the description of the data is either 
improved over or
of about the same quality as the pure hydrodynamics results.
In the second panel, our calculations are compared with the ALICE data shown
in this conference \cite{Milano:2012qm}.
It should be emphasized here that the calculation
was done {\em before} the data were shown.

For $v_2(p_T)$ in Figure \ref{fig:v2},
the $\bar{p}$ result is much improved with the afterburner at RHIC. 
This may be due to the better description of finite
baryon mean free path in UrQMD compared to hydrodynamics.
For the LHC, one may argue that 
the afterburner improves the description a little, 
but the statistical error at this point is too big to say that
definitely. Calculations with afterburner lead to larger $v_3(p_T)$ in general.
However, this needs more careful study to quantify. 

In summary, we have reported our preliminary results on coupling MUSIC to
the UrQMD afterburner. Encouragingly,
changes from the pure hydrodynamics cases seem to go in the right direction.
In near future, we plan to {\em (i)} improve our statistics up to 
10 times, {\em (ii)} include the effect of $\delta f$ at the transition,
and {\em (iii)} slightly re-tune hydrodynamics parameters such as the viscosity
for better experimental fit. These further studies will be reported
elsewhere.

\section*{Acknowledgments}

\noindent
CG, SJ and SR are supported by the Natural Sciences and
Engineering Research Council of Canada. 
BPS is supported under DOE Contract No.DE-AC02-98CH10886 and
acknowledges support from a BNL Lab Directed Research and Development grant.  
BPS gratefully acknowledges a Goldhaber Distinguished Fellowship 
from Brookhaven Science Associates.
We greatly appreciate 
computer time on the Guillimin cluster at the CLUMEQ HPC centre, a part of
Compute Canada HPC facilities where bulk of these calculations 
were carried out. 
We also thank H.~Petersen and P.~Huovinen for helpful discussions.

\section*{References}


\end{document}